\documentclass[twocolumn,showpacs,aps,prd]{revtex4}
\pdfoutput=1
\newcommand{\half}{\frac{1}{2}}

\newcommand{\frth}{\frac{1}{4}}

\newcommand{\Da}{\Delta a}

\usepackage[dvipsnames]{color}
\newcommand{\revision}[1]{{           {#1}}} 
\begin{document}
\title{Conformal theory of central surface density for
galactic dark halos}
\author{R. K. Nesbet }
\affiliation{
IBM Almaden Research Center,
650 Harry Road,
San Jose, CA 95120-6099, USA
\begin{center}rkn@earthlink.net\end{center}
}
\begin{abstract}
Numerous dark matter studies of galactic halo gravitation depend on 
models with  core radius $r_0$ and central density $\rho_0$.  Central
surface density product $\rho_0 r_0$ is found to be nearly a universal 
constant for a large range of galaxies.  Standard variational field 
theory with Weyl conformal symmetry postulated for gravitation 
and the Higgs scalar field, without dark matter, implies  nonclassical 
centripetal acceleration $\Da$, for $a=a_N+\Da$, where Newtonian 
acceleration $a_N$ is due to observable baryonic matter.  Neglecting 
a halo cutoff at very large galactic radius, conformal $\Da$ is 
constant over the  entire halo and $a=a_N+\Da$ is a universal function, 
consistent with a recent study of galaxies 
that constrains acceleration due to dark matter or to alternative 
theory.  An equivalent dark matter source is a pure cusp 
distribution with cutoff parameter determined by a halo boundary radius.
This is shown here to imply universal central surface density for any 
dark matter core model.\\
\end{abstract}
\maketitle
\section{Introduction}
Observed deviations from standard Newton/Einstein galactic gravitation 
have been modeled by distributed but unobserved dark matter (DM). 
Typical DM halo models imply centripetal radial acceleration 
$a=a_N+\Da$, a function of radius in an assumed spherical 
galactic halo. DM $\Da$ is added to baryonic Newtonian $a_N$.
\par DM core model fits to galactic rotation (orbital velocity vs. 
circular orbit radius) depend on model parameters central density 
$\rho_0$ and core radius $r_0$ for a DM halo distribution.  
Observed data imply that surface density product 
$\rho_0 r_0\simeq 100M_\odot pc^{-2}$ 
is nearly a universal constant for 
a large range of galaxies\citep{KAF04,GEN09,DON18}. 

\par Assuming Weyl conformal symmetry for field action integrals
gives an alternative explanation of observed $\Delta a$.  
Conformal gravity(CG) \citep{MAN06,MAK89,MAK94,MAN91,MAN12} and the
conformal Higgs model(CHM) \citep{NESM1,NES13,NESM2,NESM3,NESM5} 
introduce new gravitational terms in the field equations.  Current 
updated conformal theory has recently been reviewed \citep{NESM12}. 
\revision{
Nonclassical $\Da$ of spherically averaged CG
and the CHM replaces the galactic radial 
acceleration attributed to dark matter.
}

\par A recent study of rotational velocities of galaxies with 
independently measured galactic mass finds total radial acceleration 
$a$ to be a universal function of Newtonian acceleration 
$a_N$ \citep{MLS16}.  This constrains acceleration attributed to 
dark matter or to alternative theory \citep{NESM7}, requiring
$\Da$ to be a universal constant.

\revision{ 
Conformal $v^2/c^2=ra/c^2=\beta/r+\half\gamma r-\kappa r^2$ implies
$\Da=\half c^2\gamma-c^2\kappa r$, with constants 
defined by CG\citep{MAK89}. Values are fitted to observed
galactic rotation\citep{MAN06,NESM12}.
Neglecting halo cutoff $2\kappa r/\gamma$ for $r\ll r_H$ 
(halo radius), CG acceleration constant $\gamma$
predicts nonclassical acceleration $\Da$. 
$\gamma\simeq 6.35\times 10^{-28}/m$ implies
$\Da=\half\gamma c^2\simeq 0.285\times 10^{-10} m/s^2$ 
\citep{NESM3,NESM5,NESM7} for all DM core models. 
}
\par Uniform constant $\Da$ puts a severe constraint  
on any DM model.  The source density must be of the form
$\xi/r$, a pure radial cusp \citep{NESM7,NESM5}, where constant
$\xi={\Da}/{2\pi G}=0.06797kg/m^2=32.535M_\odot/pc^2$.
\revision{
CODATA Newton constant  
$G=6.67384\times 10^{-11}m^3s^{-2}kg^{-1}$ \citep{COD12}.
The conflict between cusp and core DM models \citep{deB18}, 
may rule out DM for galactic rotation.
Alternatives to CHM for Hubble expansion, introducing 
ad hoc curvature\citep{MAN06} or a cosmological constant, 
are not considered here.
}

\section{Implied DM parameters}
\par A DM galactic model equivalent to conformal theory would imply 
uniform DM radial acceleration $\Da=2\pi G\xi$, attributed to 
radial DM density $\xi/r$ for universal constant $\xi$, modified at 
large galactic radius by a halo cutoff function.
Enclosed mass $M_r=2\pi \xi r^2$ implies $r\Da/c^2=GM_r/r$. 
DM models avoid a distribution cusp by assuming finite central core 
density.
\revision{
A recent fit to Milky Way rotation uses a DM core
with decreasing exponential cutoff \citep{OU24}.
}   
For arbitrary $r_0$, asymptotic radial acceleration
is unchanged if mass within $r_0$ is redistributed to uniform 
density $\rho(r)$ within a sphere of this radius.  
Conformal density $\xi/r$ implies mass $M_0=2\pi \xi r_0^2$ in 
volume $V_0=\frac{4\pi}{3} r_0^3$. For a DM spherical model core 
that replaces a central cusp density, conformal theory implies 
constant $\rho(r_0)r_0=r_0M_0/V_0=3\xi/2$.
For assumed PI core DM density \citep{KAF04}
$\rho(r)=\rho_0 r_0^2/(r^2+r_0^2)$,
central $\rho_0=2\rho(r_0)$.  Hence for a PI core, 
\revision{
$\rho_0r_0=3\xi=\frac{3\Da}{2\pi G}
 =0.204kg/m^2= 97.6M_\odot pc^{-2}$,
independent of $r_0$.  This value is proportional to 
$\rho_0/\rho(r_0)$ for other core models.
}
DM studies indicate mean value 
$141M_\odot pc^{-2}$ \citep{DON18}. 
MOND\citep{MIL83}, assuming $a^2 \to a_N a_0$ as $a_N\to 0$,
without dark matter,
implies $\rho_0 r_0\simeq 130M_\odot pc^{-2}$ \citep{MIL09}.

\section{Conformal theory of $\Da$}
\par For a central gravitational source with spherical symmetry, in the
Schwarzschild metric, conformal gravity has an exact solution of  
radial Schwarzschild potential $B(r)$ \citep{MAK89,MAK94,MAN91}.   
Outside a source of finite radius \citep{MAK89},
\begin{eqnarray} \label{Brfn}
B(r)=-2\beta/r+\alpha+\gamma r-\kappa r^2,
\end{eqnarray}
for constants related by $\alpha^2=1-6\beta\gamma$ \citep{MAN91}. $B(r)$ 
determines circular geodesics with orbital velocity $v$ such that
$v^2/c^2=ra/c^2=\half rB^\prime(r)=\beta/r+\half\gamma r-\kappa r^2$.
The Kepler formula is $ra_N/c^2=\beta/r$, from a 2nd order equation.
The 4th order conformal equation introduces two additional constants
of motion, radial acceleration $\gamma$ and cutoff parameter $\kappa$.
Parameter $\kappa$, unique to conformal theory, relates galactic
baryonic mass to large radius $r_H$ of a galactic halo, whose cosmic 
mass has been deleted by falling into the central galaxy \citep{NESM3}.
Classical gravitation is retained at subgalactic distances by setting
$\beta=GM/c^2$ for a spherical source of baryonic mass $M$ \citep{MAN06}. 
For $r\ll r_H$, $\frac{\kappa r}{\gamma}$ can be neglected, so that
$\Da=\half\gamma c^2$.
\par The CHM \citep{NESM1,NESM7,NESM2,NESM5} determines 
$\gamma$ as a universal constant, independent of galactic mass. 
The Higgs scalar field acquires a gravitational term that implies a 
modified Friedmann equation for cosmic scale factor $s(t)$ \citep{NESM1}.  
This implies dimensionless cosmic centrifugal acceleration
$\Omega_q=\frac{s{\ddot s}}{{\dot s}^2}$. The Friedmann equation 
determines observable radial acceleration parameter $\gamma$ for 
massive objects within $r_H$ \citep{NESM5}. Assuming an empty halo, due
to gravitational concentration of all mass inside halo radius $r_H$
to within galactic radius $r_G$,  $\gamma$ is determined by 
requiring continuous acceleration across $r_H$.  Constant $\gamma$ 
has a universal value throughout a depleted halo, proportional to 
uniform cosmic mass-energy density $\rho_m$ \citep{NESM3}.  Equating 
constants for the baryonic Tully-Fisher relationship and neglecting 
cutoff $\kappa$, constant $\Delta a=\half\gamma c^2=\frth a_0$, 
for MOND $a_0$ \citep{NESM3,NESM5}. Determined by CG from observed 
galactic rotation \citep{MAN06,NESM5}, 
parameter $\gamma=6.35\times 10^{-28}/m$,
using data for the Milky Way galaxy \citep{MCG08,OAM15}.
Hence, neglecting halo cutoff, $\Delta a=0.285\times 10^{-10}m/s^2$ 
and MOND $a_0=4\Delta a=1.14\times 10^{-10}m/s^2$.

\section{Conclusions}
\par Conformal theory, consistent with the finding \citep{MLS16} for 
galaxies of known mass that observed radial acceleration $a$ is a 
universal function of baryonic $a_N$, explains the observed constancy 
of halo central surface density deduced from DM core models.
Any DM core model can be considered an approximation to implied 
conformal $\Da$. Nonclassical $\Da$ predicted by conformal theory 
would require a pure cusp mass-energy source plus halo cutoff, 
which may rule out an exact DM model. This requires 
reconsideration of the consensus LCDM paradigm. 

\vfill\eject

\begin{thebibliography}{99}  
\bibitem{KAF04} J. Kormendy and K. C. Freeman,
{\it Scaling laws for dark matter haloes in late-type and dwarf
spheroidal galaxies},
{\it Proc. IAU Symp. No. 220}, 377 (2004).                                                                                                         
\bibitem{GEN09} G. Gentile et al,
{\it Universality of galactic surface densities within one dark halo
scale-length},
{\it Nature} {\bf 461}, 627 (2009).           
\bibitem{DON18} F. Donato et al,
{\it A constant dark matter halo surface density in galaxies},
{\it MNRAS} {\bf 397}, 1169 (2018).
\bibitem{MAN06} P. D. Mannheim,
{\it Alternatives to dark matter and dark energy},
{\it Prog.Part.Nucl.Phys.} {\bf 56}, 340 (2006).
\bibitem{MAK89} P. D. Mannheim and D. Kazanas,
{\it Exact vacuum solution to conformal Weyl gravity and galactic                                                                               rotation curves},
{\it ApJ} {\bf 342}, 635 (1989).
\bibitem{MAK94} P. D. Mannheim and D. Kazanas,
{\it Newtonian limit of conformal gravity and the lack of necessity
of the second order Poisson equation},
{\it Gen.Rel.Grav.} {\bf 26}, 337 (1994).
\bibitem{MAN91} P. D. Mannheim,
{\it Some exact solutions to conformal Weyl gravity},
{\it Annals N.Y.Acad.Sci.} {\bf 631}, 194 (1991).
\bibitem{MAN12} P. D. Mannheim,  
{\it Making the case for conformal gravity},
{\it Found.Phys.} {\bf 42}, 388 (2012). 
\bibitem{NESM1} R. K. Nesbet,
{\it Cosmological implications of conformal field theory},
{\it Mod.Phys.Lett.A} {\bf 26}, 893 (2011).
\bibitem{NES13} R. K. Nesbet,
{\it Conformal gravity: dark matter and dark energy},
{\it Entropy} {\bf 15}, 162 (2013).
\bibitem{NESM2} R. K. Nesbet,
{\it Dark energy density predicted and explained},
{\it Europhys.Lett.} {\bf 125}, 19001 (2019).
\bibitem{NESM3} R. K. Nesbet,
{\it Dark galactic halos without dark matter},
{\it Europhys.Lett.} {\bf 109}, 59001 (2015).
\bibitem{NESM5} R. K. Nesbet,
{\it Conformal theory of gravitation and cosmology},
{\it Europhys.Lett.} {\bf 131}, 10002 (2020).
\bibitem{NESM12} R. K. Nesbet,
{\it Conformal theory of gravitation and cosmic expansion},
{\it MDPI Symmetry} {\bf 16}, 003 (2024).
\bibitem{MLS16} S. S. McGaugh,  F. Lelli, and J. M. Schombert,
{\it The radial acceleration relation in rotationally supported galaxies},
{\it Phys.Rev.Lett.} {\bf 117}, 201101 (2016).
\bibitem{NESM7} R. K. Nesbet,
{\it Theoretical implications of the galactic radial acceleration
relation of McGaugh, Lelli, and Schombert},
{\it MNRAS} {\bf 476}, L69 (2018).
\bibitem{COD12} P. J. Mohr, B. N. Taylor, and D. B. Newell,
{\it CODATA recommended values of the fundamental physical constants:2010},
{\it Rev.Mod.Phys.} {\bf 84}, 1527 (2012).
\bibitem{deB18} W. J. G. deBlok , 
{\it The core-cusp problem}, 
(2018), ArXiv:0910.3538v1.
\bibitem{OU24} X. Ou et al,
{\it The dark matter profile of the Milky Way inferred from its 
circular velocity curve}, 
{\it MNRAS} {\bf 528}, 693 (2024).
\bibitem{MIL83} M. Milgrom,
{\it A modification of the Newtonian dynamics: implications for galaxies},
{\it ApJ} {\bf 270}, 371 (1983).
\bibitem{MIL09} M. Milgrom,
{\it The central surface density of ‘dark halos’ predicted by MOND},
{\it MNRAS} {\bf 398}, 1023 (2009).
\bibitem{MCG08} S. S. McGaugh,
{\it Milky Way mass models and MOND},
{\it ApJ} {\bf 683}, 137 (2008).
\bibitem{OAM15} J. G. O'Brien and R. J. Moss,
{\it Rotation curve for the Milky Way galaxy in conformal gravity},
{\it J.Phys.Conf.} {\bf 615}, 012002 (2015).
\end{thebibliography}
\end{document}